\documentstyle[prl,aps,eqsecnum,epsfig,multicol]{revtex}

\begin{document}
\draft
\newcommand{\npa}[1]{Nucl.~Phys.~{A#1}}
\newcommand{\etal}{{\em et al.}}                
\title{Flow Measurements via Two-particle Azimuthal Correlations \\
       in Au~+~Au Collisions at $\sqrt{s_{NN}} = 130$ GeV.} 
\author{
K.~Adcox,$^{40}$
S.{\,}S.~Adler,$^{3}$
N.{\,}N.~Ajitanand,$^{27}$
Y.~Akiba,$^{14}$
J.~Alexander,$^{27}$
L.~Aphecetche,$^{34}$
Y.~Arai,$^{14}$
S.{\,}
% ...
\\(PHENIX Collaboration)\\
}
\address{
% ...
$^{3}$Brookhaven National Laboratory, Upton, NY 11973-5000, USA\\
% ...
$^{14}$KEK, High Energy Accelerator Research Organization, Tsukuba-shi, Ibaraki-ken 305-0801, Japan\\
% ...
$^{27}$Chemistry Department, State University of New York - Stony Brook, Stony Brook, NY 11794, USA\\
% ...
$^{34}$SUBATECH (Ecole des Mines de Nantes, IN2P3/CNRS, Universite de Nantes) BP 20722 - 44307, Nantes-cedex 3, France\\
% ...
$^{40}$Vanderbilt University, Nashville, TN 37235, USA\\
% ...
}

\date{\today}
\maketitle

\begin{abstract}

Two particle azimuthal correlation functions are presented for charged
hadrons produced in Au~+~Au collisions at RHIC
($\sqrt{s_{_{NN}}}=130$~GeV). The measurements permit determination of
elliptic flow without event-by-event estimation of the reaction plane. The
extracted elliptic flow values ($v_2$)  show significant sensitivity to
both the collision centrality and the transverse momenta of emitted
hadrons, suggesting rapid thermalization and relatively strong velocity
fields.  When scaled by the eccentricity of the collision zone
$\varepsilon$, the scaled elliptic flow shows little or no dependence on
centrality for charged hadrons with relatively low $p_T$. A breakdown of
this $\varepsilon$ scaling is observed for charged hadrons with $p_T\,>$
1.0 GeV/c for the most central collisions.

\end{abstract}
\pacs{PACS 25.75.Ld}

\begin{multicols}{2}
\narrowtext

The primary goal of current relativistic heavy ion research is the
creation and study of nuclear matter at high energy
densities\cite{lastcall-qm99,phenix-mul2001,phenix-et2001,zajcqm01,star-flow2000,Phobos-mul2000}.
Open questions include the detailed properties of such excited matter, as
well as the existence of a transition to the quark-gluon plasma (QGP)
phase.
%
%
Such a phase of deconfined quarks and gluons has been predicted to survive 
for $\approx 3-10$ fm/c in Au~+~Au collisions at the 
Relativistic Heavy Ion Collider (RHIC)\cite{zhang99}, and several experimental 
probes have been proposed for its possible detection and study\cite{lastcall-qm99}. 
Elliptic flow constitutes an important 
observable\cite{olli92,teaney2001,kolb2001,zabrodin2001,dan98,e877flow,na49-98,pin99} 
because it is thought to be driven by pressure built up early in the collision, and 
therefore can reflect conditions existing in the first few fm/c.
Elliptic flow leads to an anisotropy in the azimuthal distribution of
emitted particles. A Fourier decomposition of this
distribution\cite{gutbrod90,olli2001}
\begin{equation}
 \frac{dN}{d(\phi-\Phi_R)} \propto (1 + 
\sum_{n=1}^\infty 2 v_{n}\cos(n(\phi-\Phi_R))),
\label{vn_def}
\end{equation}
provides a characterization of the elliptic flow via the second Fourier
coefficient $v_2$. Here, $\phi$ is the azimuthal angle of an emitted
particle and $\Phi_R$ is the azimuth of the reaction plane, defined by the
beam direction and the impact parameter vector\cite{dan85}. The apparent
reaction plane is determined from the azimuthal correlations between
emitted particles, and its dispersion correction from the azimuthal
correlations between two ``subevents"\cite{dan85,olli98,poskanzer98}. 
An alternative technique for elliptic flow analysis, is the Fourier
decomposition of the pair-wise distribution in the azimuthal angle
difference ($\Delta \phi =\phi_1 - \phi_2$) between pairs of emitted
particles\cite{wang91,lacey93,laceyqm01}:
\begin{equation}
\label{cor_def}
 \frac{dN}{d\Delta \phi} \propto (1 + 
\sum_{n=1}^\infty 2 v_n^2 \cos(n \Delta \phi)) .
\end{equation}
In this case the magnitude of the elliptic flow is characterized by the
square of the second Fourier coefficient in Eq. (\ref{vn_def}), i.e.
$v_2^2$.  These methods of analysis can be taken as equivalent since (i)
the correlation between every particle and the reaction plane induces
correlations among the particles, and (ii) correlating two subevents
amounts to summing two-particle correlations\cite{olli2001}. The results
in this letter have several advantages over elliptic flow measurements
already performed at the same beam energy by the STAR\cite{star-flow2000}
and PHOBOS\cite{parkqm01} collaborations. First, two particle correlations
circumvent the need for full azimuthal detector acceptance. Second, it
allows the determination of elliptic flow without event-by-event
estimation of the reaction plane and the associated corrections for its
dispersion. Third, this method can serve to minimize many important
systematic uncertainties (detector acceptance, efficiency, etc) which may
influence the accuracy of elliptic flow measurements\cite{wang91,lacey93}.

Elliptic flow is predicted and found to be negative for beam energies
$\lesssim 4$ AGeV and positive for higher beam energies in Au~+~Au
collisions\cite{olli92,dan98,e877flow,na49-98,pin99}.  Recent theoretical
investigations have made predictions for the centrality dependence of the
scaled elliptic flow $A_2\equiv
v_2/\varepsilon$~\cite{sorge99,heiselberg99} where $\varepsilon$ is the
eccentricity or initial spatial anisotropy of the ``participant" nucleons
in the collision zone.  The eccentricity $\varepsilon$ shows an
essentially linear variation with impact parameter $b$, for $ 0.2 b_{max}
\lesssim b \lesssim 0.8 b_{max}$\cite{olli92} in Au~+~Au collisions
($b_{max}\approx 14$~fm). For central collisions ( b$\lesssim$ 5~-~6~fm),
it is predicted that higher energy densities are produced and rapid
kinetic equilibration in the QGP phase leads to a characteristic rise in
$A_2$\cite{sorge99,heiselberg99}. In addition, elliptic flow for high
$p_T$ particles has been proposed as an observable sensitive to the energy
loss of scattered partons in a QGP phase \cite{gyulassy}.  

The colliding Au beams ($\sqrt{s_{_{NN}}}=130$~GeV) used in these
measurements have been provided by the Relativistic Heavy Ion Collider at
Brookhaven National Laboratory (BNL). Charged reaction products were
detected in the east and west central arms of the PHENIX
detector\cite{phenix-mul2001,phenix-detector}. Each of these arms subtends
90$^{o}$ in azimuth $\phi$, and $\pm 0.35$ units of pseudo-rapidity
$\eta$.  The axially symmetric magnetic field of PHENIX (0.5 T) allowed
for the tracking of particles with $p_T \geq 200$ MeV/c in the fiducial
volume of both arms. The Drift Chamber (DC) and a layer of Pad Chambers
(PC1) located at radii of 2 m and 2.5 m respectively, in each arm, served
as the primary tracking detector for these measurements.  A second layer
of Pad Chambers (PC3), positioned at 5 m in the east arm, was employed to
confirm the trajectory of charged particles which traversed both the DC
and PC1.  The Zero Degree Calorimeters (ZDC), were used in conjunction
with the Beam-Beam Counters (BBC) to provide the position of the vertex
along the beam direction as well as a trigger for a wide range of
centrality selections.

The present data analysis uses two-particle azimuthal correlation
functions to measure the distribution of the azimuthal angle difference
($\Delta\phi = \phi_1 - \phi_2$)  between pairs of charged hadrons.
Following an approach commonly exploited in interferometry studies, a
two-particle azimuthal correlation function can be defined as
follows\cite{wang91,lacey93,laceyqm01}
\begin{equation}
 C(\Delta\phi) = {N_{cor}(\Delta\phi)\over{N_{uncor}(\Delta\phi)}}, 
  \label{ratio}
\end{equation}
where $N_{cor}(\Delta\phi)$ is the observed $\Delta\phi$ distribution for
charged particle pairs selected from the same event, and
$N_{uncor}(\Delta\phi)$ is the $\Delta\phi$ distribution for particle
pairs selected from mixed events. Events were selected with a collision
vertex position, $-20 <$~z~$< 20$~cm, along the beam axis. Mixed events
were obtained by randomly selecting each member of a particle pair from
different events having similar multiplicity and vertex position. In order
to suppress an over-efficiency in finding two tracks at close angles, 
hadron pairs within 1~cm of each other in the DC were removed
from both the $N_{cor}(\Delta\phi)$ and $N_{uncor}(\Delta\phi)$
distributions. Event centralities were obtained via a series of cuts in
the space of BBC versus ZDC analog response\cite{phenix-mul2001}; they
reflect percentile selections of the total interaction cross section of
6.8 barns\cite{phenix-mul2001}. Estimates for the impact-parameter and the
eccentricity, $\varepsilon \equiv\,{(<y^2>-<x^2>)}/{(<y^2>+<x^2>)}$ were
also made for each of these selections following the model detailed in
Ref.\cite{phenix-mul2001}. Here, $<\ldots>$ represents the spatial average
(weighted by the density) of participant nucleons over the transverse
plane of the collision zone\cite{kolb2001}. Systematic uncertainties
associated with the determination of $\varepsilon$ are estimated to be
$\sim$~7\%.

Correlation functions were obtained via two separate methods. In the
first, charged hadron pairs were formed by selecting each particle from a
common $p_T$ range (fixed-$p_T$ method).  In the second hadron pairs were
formed by selecting one member from a fixed $p_T$ range and the other from
outside this range (assorted-$p_T$ method).  Within statistical errors,
both methods of analysis yield similar results for the $p_T$ range
presented below, as would be expected for a system dominated by collective
motion.
      
An important prerequisite for reliable flow extraction from PHENIX data is
to establish whether or not the $\sim 180^o$ azimuthal coverage of the
detector results in significant distortions to the correlation function.
To this end, detailed simulations of the detector response, acceptance and
efficiency, have been performed for simulated data incorporating specific
amounts of flow (parameterized by $v_2$).  The results from these
simulations indicated no significant distortion to the correlation
functions due to the PHENIX acceptance. On the other hand, small
distortions to the correlation function (for $\Delta\phi \lesssim 25^o$)
as well as an incomplete recovery of $v_2$ could be attributed to
background contributions.

These background contributions principally affect the extraction of $v_2$
from the correlation function in two ways. The distortion to the
correlation function at small relative angle introduces a small systematic
distortion when fit with a Fourier function [c.f. Eq.~(\ref{cor_def})].  
A good representation of the data was obtained with the fit function
$C(\Delta \phi)= \lambda \cdot exp(-0.5(\Delta \phi/\sigma)^2) + a_1\cdot
(1+ 2v_2^2 cos(2\Delta \phi))$, where the Gaussian term is used to
characterize the background distortion at small angles.  In addition,
there is an isotropic background of false tracks which are predominantly
misidentified as high $p_T$ particles.  These contributions can be
efficiently suppressed in the east central arm of PHENIX, by requiring a
relatively stringent association between tracks found in the DC and their
associated hits in PC3.  Using such a procedure, the fraction of
background tracks has been evaluated as a function of $p_T$ and used to
correct $v_2$.  Corrections range from $\sim$ 10\% at low $p_T$ to $\sim$
25\% at 2 GeV/c with a systematic uncertainty of 5\%.  The current
analysis is restricted to the range $0.3 < p_t \le 2.5$ GeV/c to maintain
this relatively small systematic uncertainty.

\begin{figure}
\vspace*{3cm}
\centerline{\epsfig{file=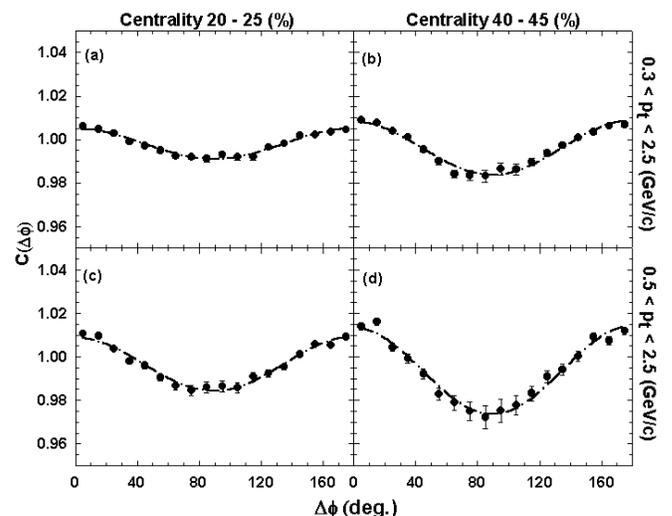,width=1.0\linewidth}}
\caption{Azimuthal correlation functions for charged hadrons as a function
of centrality and $p_T$ selection. The solid curves represent Fourier fits
following Eq. (2).  Error bars are statistical only.
}
\label{fig1}
\end{figure}

Figures \ref{fig1}a~-~d show representative $\Delta\phi$ correlation
functions obtained for charged hadrons detected in the pseudo-rapidity
range $-0.35 < \eta < 0.35$.  Correlation functions for relatively central
events (centrality = 20~-~25\%, b $\sim$ 7.0 fm)  are shown for hadrons
with $0.3 < p_T <2.5$ GeV/c and $0.5 < p_T <2.5$ GeV/c in
Figs.~\ref{fig1}a and c respectively.  The same $p_T$ selections have been
made for the correlation functions shown in Figs.~\ref{fig1}b and d but
for more peripheral collisions ( centrality = 40~-~45\%, b $\sim$ 9.6 fm)
as indicated.  Figs.~\ref{fig1}a~-~d show a clear anisotropic pattern
which is essentially symmetric about $\Delta\phi = 90^o$. There is also a
visible increase of this anisotropy with increasing impact parameter and
$p_T$. These trends are all consistent with those expected for in-plane
elliptic flow\cite{teaney2001,kolb2001,dan98,olli98}.  
     	     
\begin{figure}
\vspace*{5cm}
\centerline{\epsfig{file=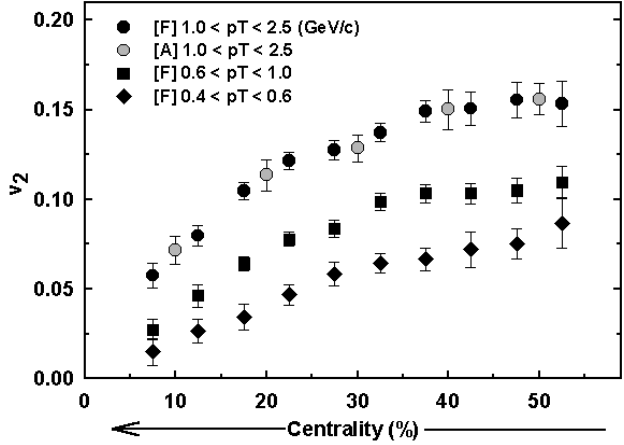,width=1.0\linewidth}}
\caption{$v_2$ vs. centrality for several $p_T$ selections. [F] and [A]
indicate results obtained with the fixed-$p_T$ and assorted-$p_T$ methods
respectively.  Systematic errors are estimated to be $\sim 5$\%; they are
dominated by the normalization of the correction function for real tracks.
For the centrality range 0-5\% the data points are statistically uncertain
and the points are omitted.}
\label{fig2}
\end{figure}

\begin{figure}
\centerline{\epsfig{file=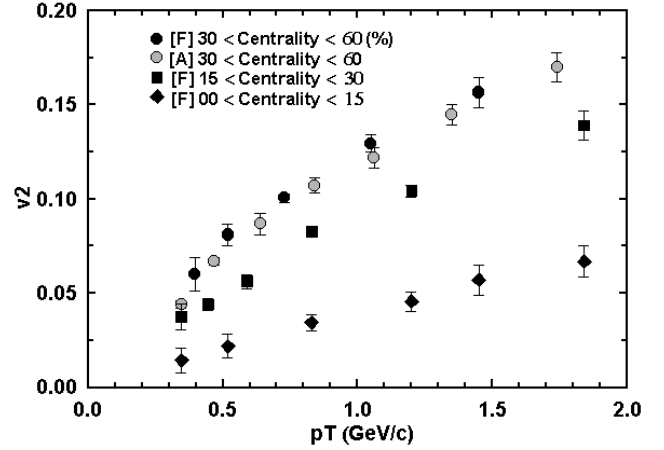,width=1.0\linewidth}}
\caption{$v_2$ vs $p_T$ for several centrality selections. [F] and [A]
follow the notation Fig.~\ref{fig2}. Systematic errors are estimated to be
$\sim 5$\%.
}
\label{fig3}
\end{figure}

The magnitude of elliptic flow and the mechanism for its development, can
be shown to be related to (a) the geometry of the collision zone, (b) the
initial baryon and energy density developed in this zone, and (c) the
detailed nature of the equation of state for the created nuclear
matter\cite{olli92,teaney2001,kolb2001,zabrodin2001,dan98,sorge99}.  
Since differential flow measurements can serve to provide important
insights for disentangling these separate
aspects\cite{teaney2001,kolb2001,zabrodin2001,dan98,sorge99}, we show the
results of such measurements in Figs.~\ref{fig2} and \ref{fig3}.
Fig.~\ref{fig2} shows $v_2$ as a function of centrality 
for several $p_T$ selections;  
$0.40 < p_T < 0.60$ (diamonds), 
$0.60 < p_T < 1.00$ (squares), and 
$1.0 < p_T < 2.5$ (circles) GeV/c respectively.  
Fig.~\ref{fig3} compares the differential flow $v_2(p_T)$, for several
centralities as indicated.
 
Figs.~\ref{fig2} and \ref{fig3} both show relatively large differential
flow values which increase with increasing impact parameter and the $p_T$
of emitted hadrons.  The separate effects of spatial asymmetry and the
response of the collision zone to the generated pressure are also
evident in Fig.~\ref{fig3}. That is, $v_2$ not only increases with
increasing impact parameter for a fixed $p_T$, but also increases with
increasing $p_T$ for a fixed centrality selection. Trivially, the
magnitude of flow should go to zero for very small and very large impact
parameters. Similarly, its magnitude can be expected to be zero for
$p_T=0$. It is interesting that Fig.~\ref{fig3} indicates an essentially
linear rise of $v_2$ with $p_T$ for each of the centrality selections
presented. Such a trend can not be accounted for via simple geometric
considerations alone\cite{huovinen2001}. However, it is compatible with
model calculations which assume a strong transverse velocity
field\cite{huovinen2001}. This suggests the presence of strong dynamically
driven transverse flow at RHIC.  
The magnitude and trends for $v_2$ shown in Figs.~\ref{fig2}
and~\ref{fig3} are consistent with other elliptic flow measurements at
RHIC\cite{star-flow2000,parkqm01}. 

\begin{figure}
\centerline{\epsfig{file=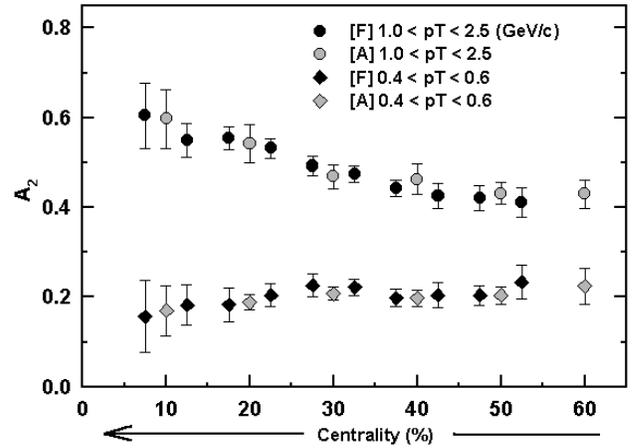,width=1.0\linewidth}}
\caption{ The centrality dependence of $A_2$ for two different $p_T$
selections. [F] and [A] follow the notation in Fig. 2. Systematic errors
are estimated to be $\sim 10$\%, dominated by the normalization of the
correction function and the model determination of$ \varepsilon$.
}
\label{fig4}
\end{figure}

Figure~\ref{fig4} aims to disentangle the geometric and dynamical ($p_T$)  
contributions to the elliptic flow over a broad range of centralities or
energy densities.  To do this, we plot $A_2$ as a function of centrality
to obtain the dynamical contributions~\cite{sorge99,heiselberg99}. This
evaluation is performed for two $p_T$ selections ($0.40 < p_T < 0.60$ and
$1.0 < p_T < 2.5$ GeV/c) which give rise to relatively low and high $p_T$
values respectively. The underlying idea is that this ratio should
remove the geometric dependence of $v_2$, while the
$p_T$ selections may provide greater sensitivity to different time scales
and energy densities associated with the expanding system.

Figure ~\ref{fig4} shows an increase in the magnitude of $A_2$ with
increasing $p_T$. This increase can be attributed to the dynamical
response of the created system, resulting from the generated pressure
gradients.  For hadrons of $0.4 < p_T <0.6$ the observed centrality
dependence of $A_2$ is compatible with $\varepsilon$ scaling. However, a
breakdown of this scaling is observed for hadrons with $1.0 < p_T < 2.5$.  
Such a trend may point to a change in the particle production mechanism or
the possibility that pressures larger than those predicted by current
hydrodynamic calculations~\cite{teaney2001,kolb2001} are being produced in
the most central collisions at RHIC. It is also interesting to note that
the species composition of the charged particle spectra changes
dramatically between the two $p_T$ ranges used in Fig.~\ref{fig4}\cite{PPG06}.


To summarize, we have measured two-particle azimuthal correlation
functions for charged hadrons produced in Au~+~Au collisions at RHIC
($\sqrt{s_{_{NN}}}=130$~GeV).  The integral, differential and scaled
elliptic flow values extracted from these measurements indicate strong
sensitivity to the collision centrality and the transverse momenta of
emitted hadrons. The centrality dependence of $v_2$ suggests that the
high-energy-density nuclear matter created at RHIC, efficiently translates
the initial spatial asymmetry into a similar asymmetry in momentum space.
The $p_T$ dependence is consistent with the development of strong
transverse velocity fields in the collision zone.  The centrality
dependence of $A_2$ for hadrons in the range $0.4 < p_T <0.6$ is
compatible with $\varepsilon$ scaling. However, a breakdown of this
scaling is observed for hadrons with $1.0 < p_T < 2.5$. Such a trend could
result from a number of effects, the most intriguing of which is a
possible change in the equation of state\cite{sorge99,heiselberg99}.
Additional experimental signatures and model calculations will undoubtedly
be necessary to test the detailed implications of these results.
Nevertheless, the results presented here clearly show that two particle
azimuthal correlation measurements provide an important probe for the
high-energy-density nuclear matter created at RHIC.


We thank the staff of the Collider-Accelerator and Physics Departments at
BNL for their vital contributions.  We acknowledge support from the
Department of Energy and NSF (U.S.A.), MEXT and JSPS (Japan), RAS,
RMAE, and RMS (Russia), BMBF, DAAD, and AvH (Germany), VR and KAW
(Sweden), MIST and NSERC (Canada), CNPq and FAPESP (Brazil), IN2P3/CNRS
(France), DAE and DST (India), KRF and CHEP (Korea), the U.S. CRDF for 
the FSU, and the US-Israel BSF.

\end{multicols}
\end{document}